\documentclass[sigconf, nonacm]{acmart}

\usepackage{enumitem}
\usepackage{multirow}
\usepackage{tcolorbox}
\tcbuselibrary{breakable}

\AtBeginDocument{%
  }

\begin{document}

\title{Is One Score Enough? Assessing Singing Quality of Songs with Temporal Score Curves}

\author{Yishan Lv$^{1,2}$, Jing Luo$^1$, Xinyu Yang$^1$, Zhizheng Wu$^2$}

\affiliation{%
  \institution{$^1$Xi'an Jiaotong University, Xi'an, China}
  \country{}
}

\affiliation{%
  \institution{$^2$The Chinese University of Hong Kong, Shenzhen, China}
  \country{}
}

\renewcommand{\shortauthors}{Lv et al.}

\begin{abstract}
Singing Quality Assessment (SQA) has become increasingly important for practical multimedia applications and Music AI systems, yet existing studies predominantly focus on short singing clips and remain insufficient for full-length songs. Unlike clip-level assessment, full-length song SQA requires modeling how singing quality varies across different audio segments and how these local variations influence the overall evaluation of vocal performance. Moreover, the scarcity of segment-level annotations makes effective supervision challenging, as directly assigning a single overall score label to every segment tends to treat different segment qualities as equivalent.
To address these challenges, we propose SongSQA, a two-stage framework for full-length song SQA. In the first stage, a Segment Score Predictor is trained with pseudo labels generated by a pre-trained teacher model, enabling segment-level singing quality prediction without requiring manual segment annotations. In the second stage, a Song Quality Aggregator integrates segment features and predicted segment scores into unified segment embeddings, and employs a learnable song embedding together with self-attention to capture the connection between segment-level vocal performance and overall song quality. In this way, SongSQA dynamically aggregates critical quality cues across the song to produce a holistic quality prediction, while also generating a temporal segment-level quality curve. Experimental results demonstrate the effectiveness of SongSQA for full-length song SQA, achieving up to a 13.95\% relative improvement in KTAU over the strongest baseline, while consistently improving other evaluation metrics across all datasets.
%

\end{abstract}

\maketitle

\section{Introduction}
The rapid expansion of online multimedia platforms and the advancement of music generative Artificial Intelligence (AI), particularly Singing Voice Synthesis (SVS)~\cite{guo2025techsinger, dai2025everyone, zhang2024stylesinger}, and Singing Voice Conversion (SVC)~\cite{zhang2025vevo2, wang2026s2voice, sha2024neural}, have led to an explosion of music content with varying quality. This surge makes manual quality assessment increasingly impractical, highlighting the urgency for automated Singing Quality Assessment (SQA) systems to handle this growing challenge. 
More broadly, in practical multimedia applications, an automated SQA system can support intelligent music education. Furthermore, in the Music AI field, it can provide valuable feedback to guide generative models toward higher generation quality.
%
However, assessing singing quality of full-length songs remains a complex human perception task. 

\begin{figure}[t]
    \centering
    \includegraphics[width=\linewidth]{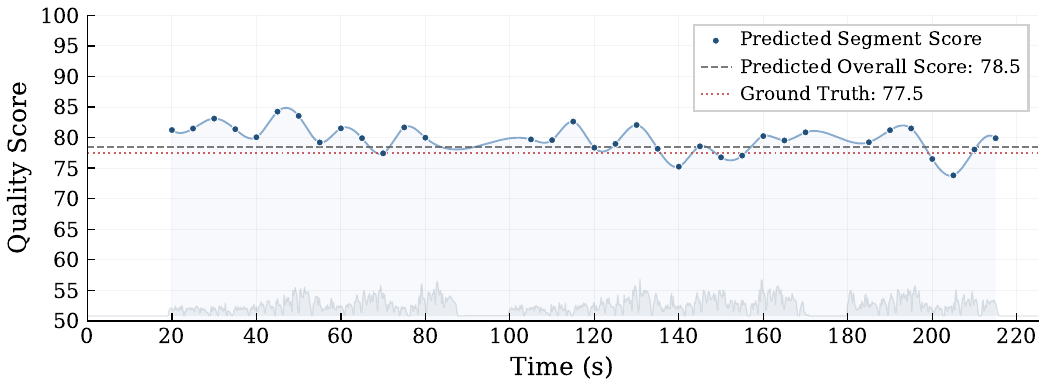}
    \caption{Temporal score curve of SongSQA on a full-length song. Beyond predicting an overall singing quality score, SongSQA predicts the quality of individual audio segments over time, forming a smoothed temporal score curve from these discrete predictions that reveals quality variations within a song that cannot be captured by a single overall score.
    } 
    \label{fig:score_curve}
\end{figure}

\begin{figure}[t]
    \centering
    \includegraphics[width=\linewidth]{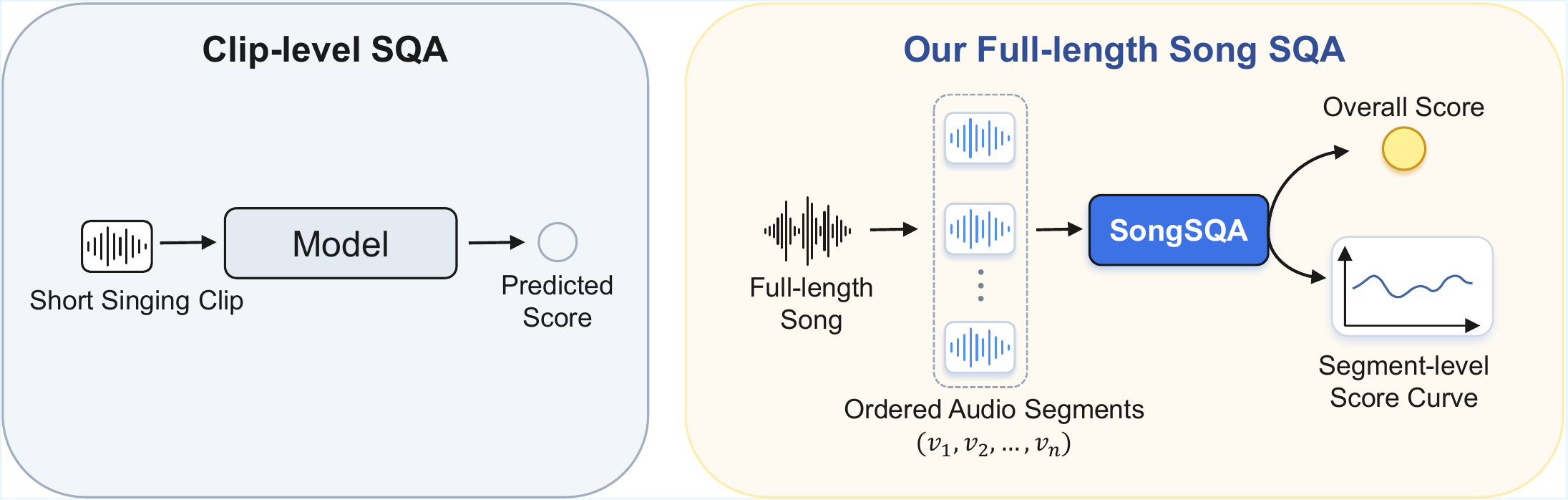}
    \caption{Comparison between conventional clip-level SQA (left) and our full-length song SQA (right). Unlike clip-level assessment that predicts a single quality score from a short singing clip, our task evaluates a complete song, producing both an overall score and a temporal segment-level quality curve.}
    \label{fig:comparison}
\end{figure}

In this work, full-length song SQA aims to evaluate the overall vocal performance of a complete song. Existing SQA studies primarily follow a clip-level paradigm, where a model takes a short singing clip as input and typically relies on pooling operations to predict a single Mean Opinion Score (MOS). While this formulation is suitable for short clips, it is insufficient for full-length songs, which should consider quality variations across different segments and their influence on the overall evaluation. Figure~\ref{fig:comparison} illustrates the difference between the conventional clip-level SQA paradigm and our full-length song SQA, where a complete song is modeled as an ordered sequence of audio segments to produce both an overall score and a temporal segment-level quality curve.

Simply transferring frameworks that are designed for SQA on singing clips, to the task of full-length song SQA faces two primary challenges. First, a full-length song is composed of diverse sections, such as verses, choruses, and bridges. Simple pooling or averaging methods often result in the loss of critical information, as they fail to distinguish between critical segments, such as expressive choruses or obvious vocal flaws, and ordinary sections, such as regular verses, while the human perception process naturally attends to specific highlights or prominent errors rather than simple averaging when evaluating the overall performance ~\cite{rozin2004feeling, zoanetti2016mitigating, gupta2018technical}. Second, the scarcity of segment-level annotations presents a major challenge for effective supervision. Most existing datasets provide only overall scores for entire songs. A common workaround is to assign the same overall score to every segment during training. However, this weak supervision strategy relies on a flawed label sharing assumption, ignoring the quality variations across different segments of a song and limiting the model's ability to capture fine-grained perceptual differences in singing quality.

To address these challenges, we propose SongSQA, a novel framework for full-length song SQA that enables fine-grained quality prediction. We design a two-stage architecture comprising a Segment Score Predictor and a Song Quality Aggregator. In the first stage, we employ a teacher model to generate pseudo labels for individual audio segments to supervise the training of the Segment Score Predictor. This enables it to extract expressive segment-level acoustic features while producing precise quality predictions. In the second stage, the Song Quality Aggregator employs the pre-trained predictor from stage I to extract segment features and segment scores, fusing them into comprehensive segment embeddings. To model the human perception process, which naturally attends to specific highlights or prominent errors, we introduce a randomly initialized learnable song embedding. Through Transformer-based self-attention, this embedding dynamically aggregates critical singing quality cues across the song to predict the final overall score. Through joint optimization, the model yields not only a single overall assessment but also a sequence of refined segment scores, thereby explicitly forming a temporal score curve.
An illustrative example of this curve is shown in Figure~\ref{fig:score_curve}.

The main contributions of this work are summarized as follows:
\begin{itemize}
    \item We propose SongSQA, a novel framework designed for full-length song SQA. Beyond predicting a single quality score, our model generates a temporal score curve to identify outstanding and underperforming segments. This approach effectively addresses the limitations of conventional paradigms, bridging the gap between conventional clip-level SQA and full-length song SQA.

    \item We design a Segment Score Predictor that employs pseudo labels generated by a teacher model. This strategy bypasses the flawed label sharing assumption caused by the scarcity of segment-level annotations, enabling the model to accurately capture precise vocal details.

    \item We introduce a Song Quality Aggregator featuring a learnable song embedding that utilizes self-attention to aggregate segment-level information. Rather than relying on simple averaging, the aggregator dynamically attends to specific expressive highlights or prominent vocal flaws, closely modeling the human perception process.
\end{itemize}


\section{Related Work}

\subsection{Early Perceptual Studies}

Singing Quality Assessment aims to computationally evaluate vocal performance in a way that approximates human perceptual judgments. In general, singing quality refers to the degree to which a vocal performance meets musical standards~\cite{gupta2018technical}. Traditional research sought to identify objective acoustic criteria that shape the subjective evaluation of singing ability. 
%
These foundational studies quantified key dimensions such as intonation and rhythmic stability~\cite{nakano2006subjective, cao2009automatic}, vibrato and diction~\cite{WAPNICK1997429}, and specific timbre characteristics~\cite{SUNDBERG2001176, cao2009automatic}. These works provided the theoretical cornerstone for early computational models in the field.

\subsection{From Reference-Based to SSL-Driven SQA}

Building upon the objective acoustic criteria, early SQA systems were predominantly formulated as reference-based evaluation frameworks. These methods evaluated a test performance by comparison with a reference recording or target melody. Lal~\cite{lal2006comparison} evaluated singing by aligning the estimated F0 of a user rendition and a reference clip. Tsai and Lee~\cite{tsai2011automatic} assessed karaoke singing using pitch, volume, and rhythm features derived from comparisons with the target performance. Molina et al.~\cite{molina2013fundamental} measured melodic similarity with respect to a target melody through F0 alignment and note-level similarity analysis. PESnQ~\cite{gupta2017perceptual} evaluated singing quality using acoustic distance features designed to reflect perceptually relevant singing parameters. However, their reliance on reference templates limits the practical applicability of such systems, since reference recordings or MIDI files are often unavailable in real world scenarios. This limitation motivated the exploration of singing evaluation without a reference.


Studies show that music experts can evaluate singers with a high level of consensus even for unfamiliar songs~\cite{nakano2006subjective}. This suggests that some aspects of human perceptual judgment can be learned directly from acoustic signals. Therefore, deep neural networks began to model singing quality from audio without relying on an explicit template. For example, Zhang et al.~\cite{zhang2019automatic} proposed a singing evaluation method without a reference melody based on a Bi-Dense neural network trained on large-scale human rating data. Huang et al.~\cite{huang2020spectral} employed a CRNN for singing quality evaluation without reference inputs and augmented spectral representations with pitch histogram features. Li et al.~\cite{li2021training} proposed an explainable singing quality assessment framework trained with augmented data, which predicts both overall singing quality and pitch accuracy. 



The emergence of Self-Supervised Learning (SSL) has extended this line of research toward more generalizable audio representations. Representative SSL models, such as wav2vec 2.0~\cite{baevski2020wav2vec} and HuBERT~\cite{hsu2021hubert}, have shown strong transferability in downstream audio prediction tasks. Cooper et al.~\cite{cooper2022generalization} demonstrated that SSL models offer stronger generalization for MOS prediction, particularly in out-of-domain settings. Inspired by this line of work, Chan et al.~\cite{chan2023improve} introduced SSL-based transfer learning into singing quality prediction, employing pre-trained speech representations to improve prediction accuracy. Bai et al.~\cite{bai2026reference} proposed a multi-feature fusion framework with integrated feature analysis for singing voice MOS prediction.

\subsection{Toward Broader Perceptual Assessment}

Recent works have further broadened automatic perceptual evaluation toward more comprehensive and unified assessment. For example, the VoiceMOS Challenge 2024~\cite{huang2024voicemos} included both synthesized speech and singing voice tracks, utilizing samples from voice synthesis and conversion systems, and SingMOS~\cite{tang2024singmos} was introduced as an open-source dataset for singing MOS prediction, while its extension SingMOS-Pro~\cite{tang2025singmospro} further enriched the benchmark with lyrics and melody annotations. Subsequently, AudioMOS Challenge 2025~\cite{huang2025audiomos} expanded the scope to synthetic audio more broadly, with MusicEval~\cite{liu2025musiceval} focusing on text-to-music clips. Jin et al.~\cite{jin2023order} evaluated music aesthetics for recommendation, and Meta Audiobox Aesthetics~\cite{tjandra2025meta} targeting unified assessment across speech, music, and sound. SongEval~\cite{yao2025songeval} further extended automatic evaluation to full-length songs, although its focus is on overall song aesthetics rather than isolated singing quality. Wang et al.~\cite{wang2025singing} targeted singing timbre popularity assessment across multiple perceptual dimensions.

Taken together, these studies highlight rapid progress in perceptual audio evaluation, while also revealing a remaining gap between short singing clips, full-song aesthetics, and the task of full-length song SQA.

\section{Method}

\subsection{Model Overview}

\begin{figure*}[t]
    \centering
    \includegraphics[width=\textwidth]{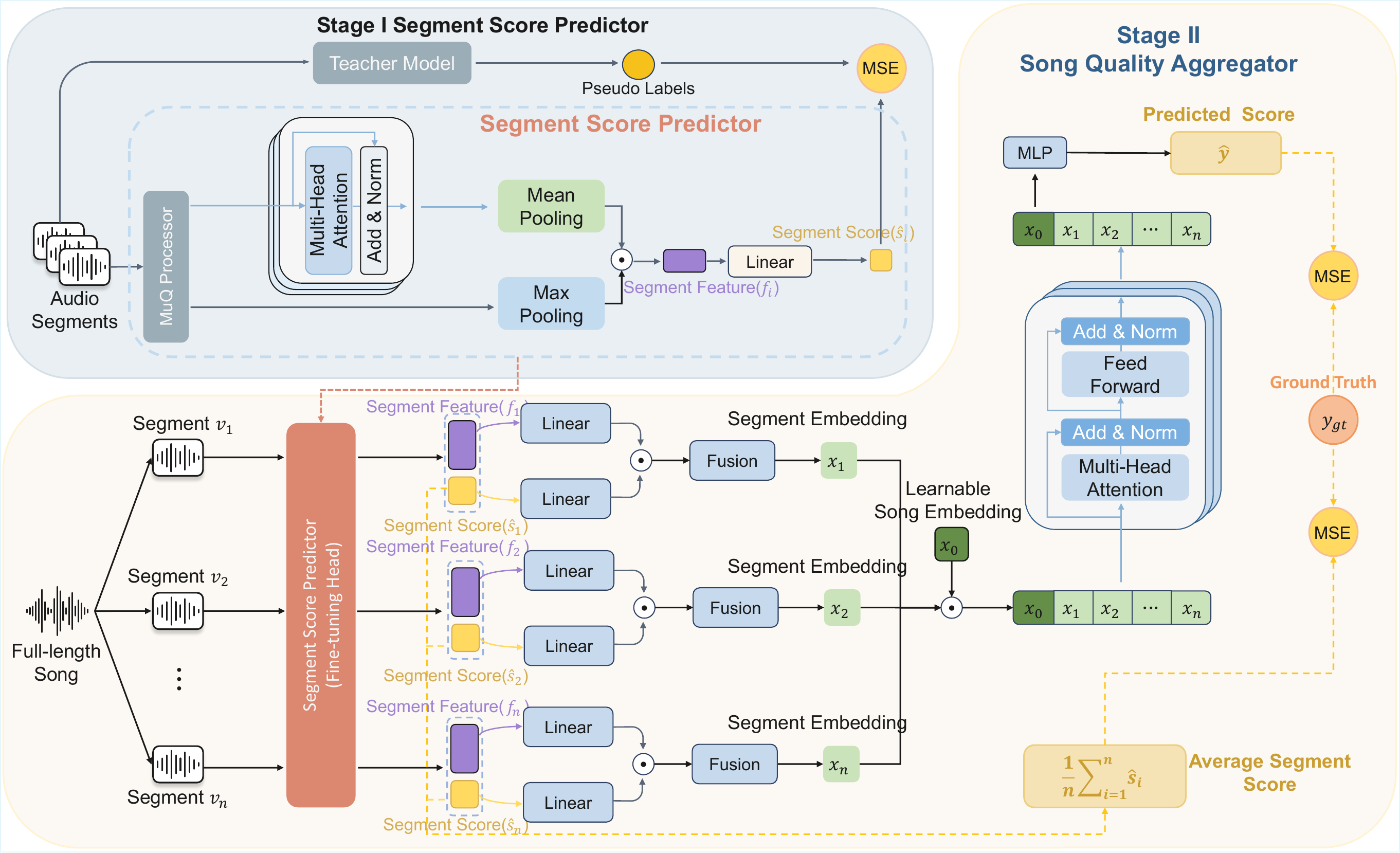}
    \caption{Model Architecture}
    \label{fig:architecture}
\end{figure*}

The proposed SongSQA framework comprises a Segment Score Predictor and a Song Quality Aggregator. Accordingly, the model employs a two-stage training strategy. The Segment Score Predictor is designed to learn segment-level quality assessment from audio segments sampled from complete songs. Specifically, music representations are extracted from the input segments using an SSL model, and pseudo labels generated by a teacher model are used for supervision, so that the Segment Score Predictor can learn fine-grained singing details. Building upon this, the Song Quality Aggregator is designed for song-level quality assessment. Given a complete song, it is first divided into ordered segments, and the Segment Score Predictor trained in the first stage is used to extract segment-level features and scores. These are further fused into segment embeddings. A randomly initialized learnable song embedding is then introduced to aggregate information from the entire sequence through the self-attention of a Transformer, enabling the prediction of the overall singing quality. The overall architecture of the proposed model is illustrated in Figure~\ref{fig:architecture}.

\subsection{Segment Score Predictor}

To address the quality variations across different audio segments within a full-length song and enable fine-grained quality perception, the first stage learns a Segment Score Predictor. By focusing on the evaluation of isolated singing segments, this stage constructs a foundational predictor with high discriminative capability for singing nuances.

Given the $i$-th audio segment $v_i$, we employ MuQ~\cite{zhu2025muq}, an SSL music representation model, to extract acoustic features. MuQ serves as a suitable backbone for our task, as it provides informative musical embeddings for perceptual singing quality.
Formally, the extracted representation sequence is denoted as
\begin{equation}
H_i^{\mathrm{muq}} = \mathrm{MuQ}(v_i).
\end{equation}
To further capture contextual dependencies within the segment, we feed $H_i^{\mathrm{muq}}$ into a self-attention encoder, yielding a representation
\begin{equation}
H_i^{\mathrm{enc}} = \mathrm{Encoder}\!\left(H_i^{\mathrm{muq}}\right),
\end{equation}
where the encoder is composed of multi-head self-attention, residual connections, and layer normalization.

Assessing singing quality is a perceptual process. In human auditory perception, the subjective evaluation of a performance is influenced by salient vocal events, ranging from technical highlights to vocal flaws. Therefore, the assessment depends not only on overall performance stability, but also on specific transient events. To capture both aspects, 
we adopt an asymmetric dual-path pooling strategy to preserve complementary perceptual cues.
Specifically, a max-pooling branch is directly applied to the original MuQ representation $H_i^{\mathrm{muq}}$ to retain prominent acoustic events. In parallel, a mean-pooling branch is applied to the contextualized representation $H_i^{\mathrm{enc}}$ to summarize the overall segment-level performance:
\begin{align}
z_i^{\mathrm{raw}} &= \mathrm{MaxPool}\!\left(H_i^{\mathrm{muq}}\right), \\
z_i^{\mathrm{ctx}} &= \mathrm{MeanPool}\!\left(H_i^{\mathrm{enc}}\right).
\end{align}
%
The final segment feature $f_i$ is obtained by concatenating these two branches,
forming a comprehensive vector that captures both prominent transient events and the average performance level derived from overall vocal stability.
%
%
\begin{equation}
f_i = \mathrm{Concat}\!\left(z_i^{\mathrm{ctx}},\, z_i^{\mathrm{raw}}\right),
\end{equation}
Subsequently, the segment feature $f_i$ is mapped through a linear layer to predict the segment score $\hat{s}_i$:
\begin{equation}
\hat{s}_i = W_r f_i + b_r.
\end{equation}

Existing datasets lack quality annotations for individual singing segments, as they usually provide only an overall score for the complete song. This makes it difficult to directly train the Segment Score Predictor. To address this, we introduce a pseudo label supervision strategy. A pre-trained teacher model is employed to evaluate the audio segments and generate pseudo labels $y_{\mathrm{pseudo}}^{(i)}$ as supervision signals.
For our teacher model, we utilize a recently proposed framework~\cite{lv2026song}, which focuses on multi-dimensional song aesthetics evaluation. To align with our objective of assessing vocal performance, we directly adopt the predicted scores from its Singing dimension branch, which evaluates timbre and vocal techniques, as pseudo labels:
%
%
\begin{equation}
y_{\mathrm{pseudo}}^{(i)} = \mathcal{T}(v_i),
\end{equation}
%
%
where $\mathcal{T}(\cdot)$ denotes the pre-trained teacher model. The training objective of Stage I is defined as
\begin{equation}
\mathcal{L}_{\mathrm{stage1}}
= \frac{1}{N}\sum_{i=1}^{N}
\left(\hat{s}_i - y_{\mathrm{pseudo}}^{(i)}\right)^2,
\end{equation}
where $N$ denotes the number of training segments. Under this pseudo label supervision, the Segment Score Predictor learns segment-level singing quality cues that are aligned with the teacher's perceptual judgments.
%

\subsection{Song Quality Aggregator}

Building upon the Segment Score Predictor, the second stage is designed to model the intrinsic connection between segment-level vocal performance and overall song quality.
Rather than relying on simple averaging, it aggregates the quality variations across different audio segments within a full-length song and produces a final song-level quality prediction.
%

Given a full-length song $V$, we divide it into an ordered sequence of audio segments, denoted as $\{v_i\}_{i=1}^{n}$. Each segment $v_i$ is then fed into the pre-trained Segment Score Predictor from Stage I. During this stage, all parameters of the predictor are frozen except for the regression head, which is further fine-tuned. Let $\Phi_{\mathrm{seg}}$ denote the operations of the Segment Score Predictor, for the $i$-th segment $v_i$, the predictor produces both a segment feature $f_i \in \mathbb{R}^{d_f}$ and a predicted segment score $\hat{s}_i \in \mathbb{R}$. 
These two outputs provide complementary cues: $f_i$ preserves acoustic details, while $\hat{s}_i$ provides a quality estimation. To unify them for song-level modeling, we first project them into distinct latent spaces and then fuse them into a quality-aware segment embedding.
%
\begin{gather}
(f_i, \hat{s}_i) = \Phi_{\mathrm{seg}}(v_i), \\
\tilde{f}_i = W_f f_i + b_f, \quad \tilde{s}_i = \sigma(W_s \hat{s}_i + b_s), \\
x_i = \mathrm{LayerNorm}\!\left( \mathrm{Concat}\!\left(\tilde{f}_i,\, \tilde{s}_i\right) \right).
\end{gather}
where $x_i \in \mathbb{R}^{d_{\mathrm{model}}}$ denotes the unified embedding of the $i$-th segment. This design provides an informative input for subsequent song-level aggregation by jointly encoding segment-level acoustic cues and score predictions.

To derive a song-level representation that reflects the overall singing quality of a full-length song, we introduce a randomly initialized Learnable Song Embedding, denoted as $x_0$. Inspired by the design of the [CLS] token in BERT~\cite{devlin2019bert}, this embedding serves as a special token for aggregating segment-level information and is prepended to the ordered segment embedding sequence.
\begin{equation}
X^{(0)} = [x_0, x_1, x_2, \dots, x_n].
\end{equation}
The sequence is processed by an aggregator, denoted as $\mathrm{Agg}$, which is composed of multiple Transformer encoder layers.
\begin{equation}
X^{(\ell)} = \mathrm{Agg}^{(\ell)}\!\left(X^{(\ell-1)}\right), \quad \ell = 1,2,\dots,L,
\end{equation}
%
%
%
where $L$ is the number of layers.
Through self-attention, the Learnable Song Embedding $x_0$ dynamically interacts with all segment embeddings $x_i$. 
This enables the model to adaptively emphasize segments that are more influential to the overall song quality, instead of treating all segments equally.
%
As the sequence passes through the encoder layers, critical segment-level information is aggregated into the special token at the beginning of the sequence. We then extract the output at this position as the song-level representation, denoted as $h_{\mathrm{global}}$.
\begin{equation}
h_{\mathrm{global}} = X^{(L)}_0.
\end{equation}
An MLP regression head is then used to predict the final singing quality score of the full-length song:
\begin{equation}
\hat{y} = \mathrm{MLP}\!\left(h_{\mathrm{global}}\right).
\end{equation}
%
During the training phase of Stage II, optimizing only the song-level regression objective may cause the segment scores to drift away from the segment-level semantics learned in Stage I. To mitigate this issue, we introduce a segment-score consistency regularization, which encourages the average predicted segment score to remain compatible with the song-level supervision.
Specifically, the consistency regularization is formulated as
\begin{equation}
\mathcal{L}_{\mathrm{cons}}
= \left(\frac{1}{n}\sum_{i=1}^{n}\hat{s}_i - y_{\mathrm{gt}}\right)^2,
\end{equation}
while the song-level regression loss is defined as
\begin{equation}
\mathcal{L}_{\mathrm{song}} = \left(\hat{y} - y_{\mathrm{gt}}\right)^2.
\end{equation}
The overall training objective of Stage II is
\begin{equation}
\mathcal{L}_{\mathrm{stage2}}
= \mathcal{L}_{\mathrm{song}} + \lambda \mathcal{L}_{\mathrm{cons}},
\end{equation}
where $\lambda$ is a balancing coefficient. Consequently, this formulation preserves the discriminative capability learned at the segment level while enabling adaptive song-level aggregation over diverse local singing qualities.


\section{Experiment}

\subsection{Experiment Setup}
\label{sec:setup}

\textbf{Datasets.}
To comprehensively evaluate the effectiveness and generalization capability of our proposed model, we conduct experiments on two singing quality assessment datasets: a widely adopted public benchmark annotated by non-expert listeners, the Lyra Lab Singing Assessment (Lyra-SA) Dataset, and a proprietary dataset rigorously evaluated by professional music experts, our Internal Dataset.

Lyra-SA Dataset is a publicly available dataset for singing quality assessment, released by Tencent Music Lyra Lab.\footnote{\url{https://lyracobar.y.qq.com/singvoicedataset.html}} All recordings in this dataset were collected from the WeSing platform, a large-scale online karaoke platform in China. In total, Lyra-SA contains 1,000 full-length singing recordings, covering 10 songs with 100 user covers for each song. The dataset was collected from a diverse set of users, including 516 female and 484 male singers, without explicit restrictions on gender, age, or region. To improve diversity and representativeness, multiple samples from the same singer were strictly excluded. Each sample was subjectively rated by non-expert listeners and assigned an overall perceptual score ranging from 0 to 1.

Internal Dataset is a proprietary dataset for singing quality assessment. It contains 2,035 audio recordings covering 1,284 different songs, with isolated vocal tracks extracted via the MDX-Net~\cite{kim2021kuielab} source separation model. The dataset comprises performances from 1,250 singers (755 female and 495 male). In terms of musical style, it encompasses a variety of popular genres, such as folk, ballad, R\&B, and rock. The singing language is primarily Mandarin and Cantonese, with a small proportion of Hokkien. To ensure highly reliable annotations, the dataset was evaluated by four professional music experts. When assigning the overall singing quality score (on a 0--100 scale), the experts considered eight vocal attributes: pitch and rhythm, timbre and distinctiveness, vocal range, diction and articulation, fundamental vocal techniques, stylistic nuances, narrative delivery, and emotional expressiveness. Although these specific attributes were also rated on a 1--5 scale, we exclusively utilize the overall score as the label during model training.

\textbf{Baseline Models.}
To validate the effectiveness of our proposed model on the singing quality assessment task, we select several representative state-of-the-art (SOTA) models for comparison. For a fair evaluation, all baselines are retrained on our training sets under identical experimental settings.
\begin{itemize}[leftmargin=*]
    \item \textbf{UTMOSv2} ~\cite{baba2024t05}: A top-performing model from the VoiceMOS challenge 2024~\cite{huang2024voicemos}. It employs a multi-feature fusion strategy, concatenating self-supervised representations from wav2vec 2.0 with mel-spectrogram features extracted via a pre-trained image classifier, followed by a fully connected layer for quality prediction on singing clips.
    \item \textbf{PS-SQA} \cite{shi2024pitch}: Another top-tier model from the VoiceMOS challenge 2024~\cite{huang2024voicemos}. It combines a self-supervised learning framework with pitch histograms and non-quantized spectrogram features. This model introduces a bias correction branch and an ensembling strategy to enhance clip-level prediction accuracy.
    \item \textbf{RAMP+} \cite{wang2025ramp+}: A model designed for assessing singing clips. It dynamically integrates scores derived from self-supervised acoustic features with feature retrieval-based scores to produce a comprehensive quality rating.

    \item \textbf{SSL-based Frameworks}: To evaluate the impact of different pre-trained SSL representations, we adapt a simple SSL fine-tuning paradigm for full-length song SQA, following Cooper et al. \cite{cooper2022generalization}. Specifically, this paradigm applies mean pooling over the output embeddings of a pre-trained backbone, followed by a linear regression head for score prediction. In our experiments, we instantiate this framework with three SSL backbones: wav2vec 2.0 \cite{baevski2020wav2vec}, which provides general audio representations, and MERT \cite{li2024mert} and MuQ \cite{zhu2025muq}, both of which are designed for music understanding. Accordingly, we denote the resulting models as \textbf{wav2vec 2.0-based}, \textbf{MERT-based}, and \textbf{MuQ-based} in our subsequent evaluations.
    
\end{itemize}

\begin{table*}[t]
\centering
\caption{Comparison results of SongSQA and baseline models on the Lyra-SA and Internal datasets. UTMOSv2, PS-SQA, and RAMP+ are representative SOTA methods for clip-level SQA, while the three SSL-based baselines are our full-length song adaptation paradigms built upon pre-trained audio representations. All scores in both datasets are normalized to a 0--100 scale. $\uparrow$ ($\downarrow$) indicates that a higher (lower) value is better. Best results are highlighted in \textbf{bold}.}
\label{tab:main_results}
\begin{tabular*}{\textwidth}{@{\extracolsep{\fill}}lcccc|cccc@{}}
\toprule
\multirow{2}{*}{\textbf{Model}} & \multicolumn{4}{c}{\textbf{Lyra-SA Dataset}} & \multicolumn{4}{c}{\textbf{Internal Dataset}} \\
\cmidrule{2-5} \cmidrule{6-9}
& \textbf{MSE $\downarrow$} & \textbf{LCC $\uparrow$} & \textbf{SRCC $\uparrow$} & \textbf{KTAU $\uparrow$} & \textbf{MSE $\downarrow$} & \textbf{LCC $\uparrow$} & \textbf{SRCC $\uparrow$} & \textbf{KTAU $\uparrow$} \\
\midrule
UTMOSv2 & 193.55 & 0.5312 & 0.5520 & 0.3956 & 144.52 & 0.8133 & 0.7892 & 0.5901 \\
PS-SQA & 147.55 & 0.5247 & 0.5185 & 0.3755 & 73.29 & 0.8246 & 0.8014 & 0.6011 \\
RAMP+ & 149.41 & 0.4804 & 0.4679 & 0.3354 & 83.17 & 0.8032 & 0.7952 & 0.6000 \\
\midrule
wav2vec 2.0-based & 151.79 & 0.4915 & 0.4899 & 0.3531 & 79.40 & 0.8011 & 0.7883 & 0.5884 \\
MERT-based & 163.45 & 0.4515 & 0.4386 & 0.3104 & 79.99 & 0.7988 & 0.7886 & 0.5952 \\
MuQ-based & 131.64 & 0.5842 & 0.5800 & 0.4180 & 52.61 & 0.8814 & 0.8106 & 0.6249 \\
\midrule
\textbf{SongSQA (Ours)} & \textbf{114.01} & \textbf{0.6237} & \textbf{0.6317} & \textbf{0.4662} & \textbf{37.56} & \textbf{0.9235} & \textbf{0.8859} & \textbf{0.7121} \\
\bottomrule
\end{tabular*}
\end{table*}

\textbf{Evaluation Metrics.}
To evaluate model performance, we employ four standard metrics commonly adopted in prior quality assessment work \cite{huang2024voicemos, huang2025audiomos}, covering both prediction accuracy and rank consistency between the model's predictions and ground truth scores. Specifically: (1) \textbf{Mean Squared Error (MSE)} computes the average squared difference between predicted and actual scores, where a lower value signifies superior absolute accuracy; (2) \textbf{Linear Correlation Coefficient (LCC)} measures the degree of linear dependence, with higher values indicating better trend alignment; (3) \textbf{Spearman Rank Correlation Coefficient (SRCC)} evaluates the consistency of the predicted ranking with the human-annotated ranking, and (4) \textbf{Kendall's Tau (KTAU)} similarly assesses rank correlation but exhibits stronger robustness when handling tied ranks. For SRCC and KTAU, values closer to 1 indicate stronger ranking capability.

\textbf{Implementation Details.}
For both the Lyra-SA and Internal Dataset, we randomly split the data into training, validation, and test sets with a ratio of 8:1:1. All ground truth scores are normalized to a 0--100 scale. The models are trained on the training set, and the best checkpoint is selected on the validation set. Final evaluation is conducted on the test set. To reduce the influence of data partition randomness, we perform five independent random splits for each dataset and report the arithmetic mean of the results across the five runs.

Since Stage I of SongSQA focuses on segment-level singing quality prediction, each song is first partitioned into audio segments of 10 seconds, with an overlap of 5 seconds during preprocessing. Segments consisting primarily of silence are filtered out. During Stage I training, all extracted audio segments are randomly shuffled across the training set to optimize the Segment Score Predictor. In Stage II, each song is segmented into an ordered sequence of audio segments, which is then used as input to the Song Quality Aggregator for final song-level quality prediction.

In Stage I, the Segment Score Predictor is built with two self-attention layers. In Stage II, the number of layers $L$ in the Song Quality Aggregator is set to 2. The loss weight coefficient $\lambda$ is set to 0.1. All models are optimized using Adam, with $\beta_1 = 0.9$, $\beta_2 = 0.999$, and $\epsilon = 10^{-8}$.

\subsection{Main Results}

To evaluate the effectiveness of the proposed SongSQA framework for singing quality assessment, we conduct comparative experiments against several SOTA methods introduced in Section~\ref{sec:setup}, including UTMOSv2, PS-SQA, RAMP+, as well as SSL-based baselines built on wav2vec 2.0, MERT, and MuQ. The experiments are performed on both the Lyra-SA dataset and the Internal Dataset. The results on the test set and their detailed analysis are presented below.

\begin{table*}[ht]
\centering
%
\caption{Ablation study on three variant groups: Teacher Models, Feature Fusion Strategy, and Aggregation Design. All scores in both datasets are normalized to a 0--100 scale. $\uparrow$ ($\downarrow$) indicates that a higher (lower) value is better. The best results are highlighted in \textbf{bold}, and the second-best results are \underline{underlined}.}
\label{tab:ablation_results}
\begin{tabular*}{\textwidth}{@{\extracolsep{\fill}}lcccc|cccc@{}}
\toprule
\multirow{2}{*}{\textbf{Variant}} & \multicolumn{4}{c}{\textbf{Lyra-SA Dataset}} & \multicolumn{4}{c}{\textbf{Internal Dataset}} \\
\cmidrule{2-5} \cmidrule{6-9}
& \textbf{MSE $\downarrow$} & \textbf{LCC $\uparrow$} & \textbf{SRCC $\uparrow$} & \textbf{KTAU $\uparrow$} & \textbf{MSE $\downarrow$} & \textbf{LCC $\uparrow$} & \textbf{SRCC $\uparrow$} & \textbf{KTAU $\uparrow$} \\
\midrule
w/ SongEval & 136.01 & 0.5509 & 0.5392 & 0.3921 & 66.79 & 0.8296 & 0.6191 & 0.4434 \\
w/ Gemini-3-Pro & 127.16 & 0.5808 & 0.5873 & 0.4261 & 44.13 & 0.8912 & 0.8325 & 0.6510 \\
w/ Ground Truth & 118.02 & 0.6115 & 0.6128 & 0.4549 & 37.20 & 0.9084 & 0.8701 & 0.6911 \\
\midrule
w/o Scores & 115.32 & \textbf{0.6277} & 0.6284 & \textbf{0.4681} & \textbf{35.51} & 0.9230 & 0.8837 & 0.7107 \\
w/o Features & 187.55 & 0.5315 & 0.5152 & 0.3765 & 200.32 & 0.8221 & 0.8307 & 0.6474 \\
\midrule
w/o SongEmb & 125.54 & 0.6042 & 0.6083 & 0.4443 & 39.44 & 0.9097 & 0.8785 & 0.7043 \\
w/o SongEmb \& TE & 117.88 & 0.6158 & 0.6110 & 0.4501 & \underline{35.71} & 0.9224 & 0.8794 & 0.7049 \\
\midrule
\textbf{SongSQA} & \textbf{114.01} & \underline{0.6237} & \textbf{0.6317} & \underline{0.4662} & 37.56 & \textbf{0.9235} & \textbf{0.8859} & \textbf{0.7121} \\
\bottomrule
\end{tabular*}
\end{table*}

\textbf{Results on Lyra-SA Dataset.}
Table~\ref{tab:main_results} shows the comparison results on the Lyra-SA dataset. 
SongSQA obtains an MSE of 114.01, achieving a considerably lower prediction error than the strongest baseline, MuQ-based, with an MSE of 131.64. It also consistently outperforms other baseline methods, such as UTMOSv2, PS-SQA, RAMP+, and the wav2vec 2.0-based and MERT-based variants, demonstrating stronger accuracy in predicting absolute quality scores. In terms of ranking consistency, SongSQA achieves an SRCC of 0.6317 and a KTAU of 0.4662, while the best-performing baseline model only reaches an SRCC of 0.5800 and a KTAU of 0.4180, highlighting the capability of our framework in preserving the relative quality ranking among different songs. SongSQA further attains the highest LCC of 0.6237, which reflects a stronger linear correlation with human subjective ratings.
Overall, by achieving the best performance across all four evaluation metrics, these results comprehensively demonstrate the effectiveness of the proposed SongSQA framework compared to existing models.

A closer comparison with the baseline models further reveals that the methods originally developed for assessing short singing or audio clips, such as UTMOSv2, PS-SQA, and RAMP+, remain clearly limited when applied to full-length song SQA. They do not show clear advantages over basic full song variants, such as the wav2vec 2.0-based and MERT-based models, and fall noticeably behind the strongest MuQ-based baseline. This suggests that full-length song SQA cannot be determined solely from isolated local cues, but instead requires a more holistic judgment over quality variations across different segments of the song. Therefore, simple feature regression or aggregation is insufficient for this task, particularly when the target score is formed through human perception over an entire musical production.
Notably, SongSQA clearly outperforms the MuQ-based baseline using the same underlying MuQ representations. This indicates that the improvement does not simply come from a stronger backbone, but from the proposed two-stage design. By first capturing segment-level details through the Segment Score Predictor, and then integrating them via the Song Quality Aggregator, SongSQA demonstrates strong capability for full-length song SQA.

\textbf{Results on Internal Dataset.}
Table~\ref{tab:main_results} also reports the comparison results on the Internal Dataset. The overall findings are consistent with those observed on Lyra-SA. SongSQA again achieves the best performance across all four evaluation metrics, further confirming the effectiveness of the proposed framework for full-length song SQA. 


Compared with the results on Lyra-SA, all methods achieve better results on the Internal Dataset. One possible reason is that the Internal Dataset is annotated by professional music experts, thus providing more consistent supervision, whereas Lyra-SA is rated by non-expert listeners and may therefore involve greater subjective variability. This comparison suggests that, although full-length song SQA is challenging in both settings, more stable supervision allows the proposed framework to better demonstrate its capability in assessing singing quality.

\subsection{Ablation Study}


To further examine the contribution of the key components in SongSQA, we conduct ablation studies on both the Lyra-SA dataset and the Internal Dataset. Specifically, the ablations are designed from three perspectives.


\textbf{Ablation on Teacher Models.}
To investigate the effect of pseudo label quality in Stage I, we conduct an ablation study on the source of pseudo labels. We construct three variants by replacing the teacher model used to generate pseudo labels for each audio segment:

\begin{itemize}[leftmargin=*]
    \item \textbf{w/ SongEval}:
    SongEval~\cite{yao2025songeval} is a benchmark developed for song aesthetics evaluation. We utilize the authors' publicly released checkpoint to perform inference on audio segments, and adopt the predicted scores from its \textit{Naturalness} branch, which focuses on vocal performance, as pseudo labels.

    \item \textbf{w/ Gemini-3-Pro}:
    Gemini-3-Pro is a large multimodal model with strong capabilities in audio understanding. In this variant, we prompt the model to evaluate the singing quality of each audio segment and output a numerical score, which is then used as the corresponding pseudo label. The prompt is given as follows:

    \begin{tcolorbox}[breakable, colback=gray!5!white, colframe=gray!50!black, arc=2mm, boxrule=0.5pt, left=2mm, right=2mm, top=2mm, bottom=2mm]
    \small
    \textbf{Task:} Conduct a fine-grained Singing Quality Assessment (SQA) on the audio. Analyze the acoustic and musical characteristics based on professional music production standards.
    
    \textbf{Evaluation Scale:} \\
    Utilize a continuous 0--100 scalar metric. \\
    90--100 (Professional): Studio quality, flawless pitch, professional technique. \\
    80--89 (Good): Solid performance, minor flaws but pleasant. High-level amateur. \\
    70--79 (Average): Noticeable pitch/timing issues, average amateur level. \\
    60--69 (Poor): Frequent off-key notes, unstable rhythm, or bad recording quality. \\
    $<$ 60 (Bad): Unusable audio, severe distortion, or completely out of tune.
    
    \textbf{Evaluation Dimensions:} \\
    Overall Score: Overall satisfaction, listenability, performance maturity.
    
    \textbf{Output JSON ONLY:} \\
    \{"overall\_score": int\}
    \end{tcolorbox}

    \item \textbf{w/ Ground Truth}: This variant is designed to examine the effect of removing inference from the teacher model, and instead directly assigning the overall ground truth score of the full-length song to all of its audio segments as pseudo labels. As a result, all segments from the same song share an identical training target. This setting allows us to test the flawed label sharing assumption, demonstrating how ignoring the inherent quality variations across different segments limits the capability of the model to capture perceptual differences in singing quality.

\end{itemize}

Table~\ref{tab:ablation_results} presents the ablation results on different sources of pseudo labels. On the Lyra-SA dataset, the full SongSQA model achieves the best performance across all metrics. Notably, the \textbf{w/ Ground Truth} variant, which directly assigns the overall score to all segments, still performs worse than the full model (e.g., 118.02 vs. 114.01 in MSE). This result highlights the limitation of label sharing across segments, as it overlooks quality variations within a song. In contrast, the generated pseudo labels provide more suitable supervision, enabling the model to learn more discriminative segment-level acoustic cues.

%
A similar trend is observed on the Internal dataset. Compared with Lyra-SA, whose annotations are more affected by subjective variation from non-expert listeners, the Internal dataset is annotated by experts and therefore provides more consistent supervision. In this setting, although the \textbf{w/ Ground Truth} variant achieves a slightly lower MSE, it performs worse on the ranking-related metrics, while the full model remains superior overall. This suggests that sharing the overall score across all segments may make it easier for the model to fit an average score, but weakens its ability to capture quality differences across segments.

%
Regarding the alternative teacher models, \textbf{w/ SongEval} performs the worst on both datasets. A possible reason is that its \textit{Naturalness} branch mainly emphasizes phrasing quality and breath control, rather than overall singing quality. \textbf{w/ Gemini-3-Pro} performs better than SongEval, suggesting that large multimodal models can provide useful segment-level evaluations. However, it still underperforms the full model, indicating that it is less effective at capturing subtle perceptual cues related to singing quality. These results suggest that the effectiveness of SongSQA depends not only on the availability of segment-level supervision, but also on how well the generated pseudo labels align with singing quality.

\textbf{Ablation on Feature Fusion Strategy.}
To evaluate the contributions of the segment features and the segment scores output by Segment Score Predictor, we design two ablation variants regarding the feature fusion strategy in Song Quality Aggregator:

\begin{itemize}[leftmargin=*]
    \item \textbf{w/o Scores}: Removes the segment score branch, feeding only the extracted segment features into the Song Quality Aggregator.
    
    \item \textbf{w/o Features}: Removes the segment feature branch, feeding only the predicted segment scores into the Song Quality Aggregator.
\end{itemize}

Table~\ref{tab:ablation_results} further reports the ablation results on the feature fusion strategy. Across both datasets, the \textbf{w/o Features} variant performs substantially worse than the other variants, showing that predicted segment scores alone are insufficient for reliable song-level singing quality assessment. This suggests that segment features remain the main source of perceptually relevant singing information. Meanwhile, the comparison between the full model and \textbf{w/o Scores} shows that features alone are also not optimal. Although \textbf{w/o Scores} achieves slightly better results on a few individual metrics, the full SongSQA model still achieves better overall performance, outperforming it on five of the eight metrics across the two datasets. 
These results indicate that the predicted segment scores provide complementary quality cues beyond the segment features, and that combining both yields more informative inputs for Song Quality Aggregator, leading to better overall singing quality prediction.
%

%

\textbf{Ablation on Aggregation Design.}
To verify the effectiveness of the aggregation design in Song Quality Aggregator, which utilizes a learnable song embedding $x_0$ and a Transformer Encoder to aggregate segment-level information, we design two architectural ablation variants:

\begin{itemize}[leftmargin=*]
    \item \textbf{w/o SongEmb}: Removes the learnable song embedding $x_0$. Feeds only the sequence of segment embeddings $[x_1, x_2, \dots, x_n]$ into the Transformer Encoder to model contextual dependencies, aggregates the outputs via mean pooling, and passes them to the MLP for the final score prediction.
    
    \item \textbf{w/o SongEmb \& TE}: Further removes the Transformer Encoder, thereby eliminating contextual interactions among segments. Directly applies mean pooling to the isolated segment embeddings to derive the overall song representation, which is then fed into the MLP for score regression.
\end{itemize}

Table~\ref{tab:ablation_results} reports the ablation results regarding the aggregation design in the Song Quality Aggregator. On the Lyra-SA dataset, the full model achieves the best overall performance. This suggests that directly applying mean pooling to segment embeddings is insufficient, as it tends to smooth out quality variations across segments. In contrast, the full model introduces a learnable song embedding to aggregate segment-level information adaptively, allowing the model to place more emphasis on segments that are more informative for overall singing quality. The \textbf{w/o SongEmb} variant further shows that contextual modeling by the Transformer Encoder alone is not sufficient. Without an explicit song-level token to summarize segment interactions, the contextual information modeled by the Transformer Encoder is still reduced by mean pooling, limiting the effectiveness of aggregation.

A similar pattern can be observed on the Internal dataset. Although the \textbf{w/o SongEmb \& TE} variant achieves a slightly lower MSE, the full model still performs best on the ranking-related metrics. This suggests that, under relatively more consistent expert annotations, simple mean pooling can fit the average song-level score more easily, but is less effective at preserving relative quality differences across songs. Moreover, the \textbf{w/o SongEmb} variant remains inferior even to \textbf{w/o SongEmb \& TE}, which indicates that contextual modeling alone does not necessarily improve song-level assessment when there is no effective way to summarize segment interactions into a song-level representation. In summary, these results show that the learnable song embedding plays a central role in SongSQA, and that the Transformer Encoder becomes more effective when coupled with it for song-level aggregation.

\subsection{Case Study}
\label{case study}

We conduct a detailed case study to demonstrate how our model captures local variations in singing quality. Specifically, the case study analyzes a representative song to illustrate these fine-grained predictions.

\begin{figure}[t]
    \centering
    \includegraphics[width=\linewidth]{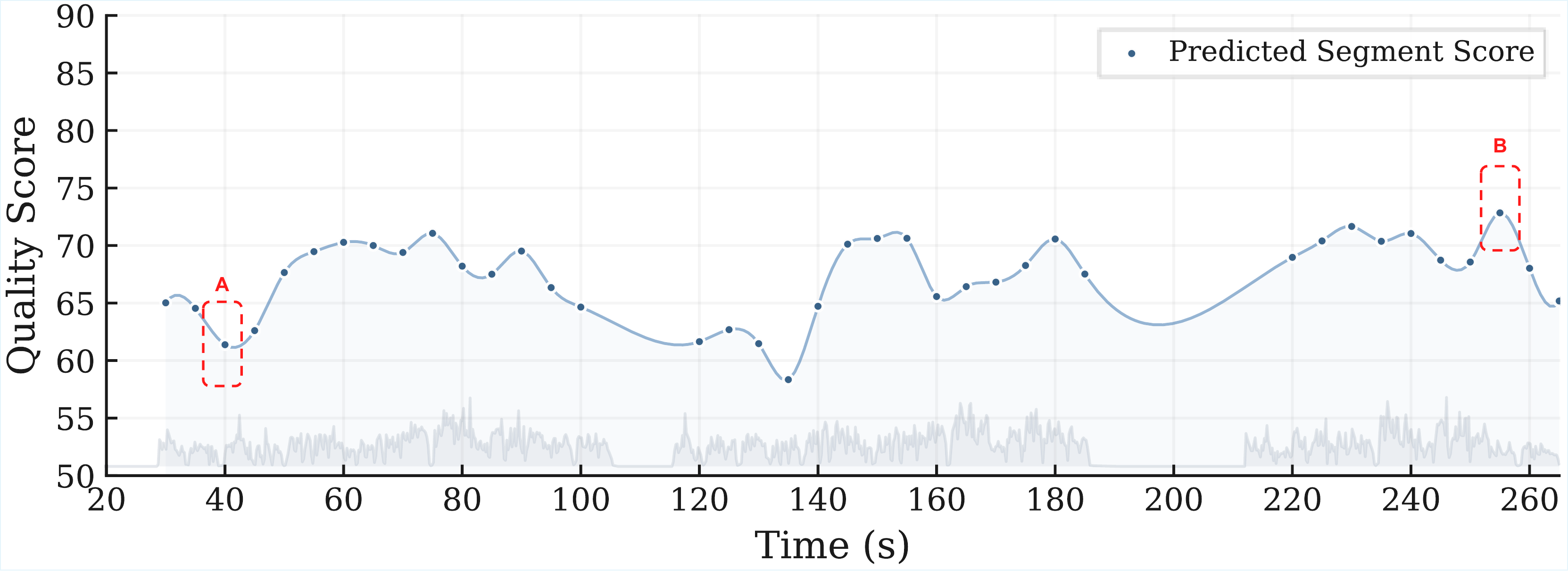}
    \caption{Predicted temporal score curve generated by SongSQA for a complete song. The red dashed boxes highlight Segment A and Segment B, which represent the lower scoring and higher scoring excerpts of Song 1 provided in our demo page and analyzed in Section~\ref{case study}.
    } 
    \label{fig:case_study}
\end{figure}

Fig.~\ref{fig:case_study} shows the temporal score curve predicted by SongSQA for a representative song from the test set. Beyond predicting a single overall quality score, SongSQA reveals clear quality variations across individual segments, reflecting the dynamic nature of vocal performance throughout a complete track. To further examine whether these local variations are perceptually meaningful, we highlight two representative regions for detailed analysis. These regions correspond directly to the two excerpts of Song 1 provided on our demo page: Segment A, representing a lower predicted quality, and Segment B, representing a higher predicted quality.

In the segment receiving a lower score, the vocal delivery appears less stable, particularly in phrases within the lower vocal register, where the descending pitches are not fully supported, resulting in an unstable tonal center. This leads to a weaker sense of intonational precision and phrase support. In addition, the phrasing is comparatively less cohesive, with several note transitions sounding less smoothly connected, which reduces the overall fluency of the performance. By contrast, the segment achieving a higher score exhibits a more stable pitch contour and a clearer tonal center, especially during sustained notes and phrase transitions. The vocal line appears better supported, with more consistent breath control and a more continuous phrase shape, resulting in a stronger sense of projection and musical coherence.


Taken together, the comparison suggests that the predicted score variations align closely with human perception. As demonstrated in this case study, the model successfully tracks local variations in vocal delivery that influence human evaluation.

\subsection{Subjective Evaluation}

To further examine whether the segment-level quality differences identified by SongSQA are perceptually meaningful, we conducted a small-scale pairwise subjective evaluation. We randomly selected five full-length songs from the test set. For each song, we selected two audio segments according to the predictions of our proposed framework, SongSQA, including one segment predicted as lower quality (Segment A) and one predicted as higher quality (Segment B). Participants were asked to conduct a blind pairwise evaluation and identify the segment exhibiting better singing performance.

A total of 12 listeners participated in the evaluation, including 8 males and 4 females. To account for varying levels of musical expertise, the participant pool included 8 amateur music enthusiasts, and 4 individuals with systematic musical training or professional musical backgrounds.

Across the five songs, 52 out of 60 valid judgments were aligned with the model predictions, corresponding to an agreement rate of 86.7\%. As summarized in Table~\ref{tab:subjective}, the agreement remained consistently high across most song pairs. These results suggest that the segment-level quality differences identified by SongSQA are generally consistent with human perception. Given the modest scale of this study, this subjective evaluation serves as further perceptual evidence supporting the effectiveness of the proposed framework.

For further auditory reference, the audio samples of the selected segments discussed in both the case study and the subjective evaluation have been made available online.\footnote{\url{https://anon-music.github.io/audio-eval-demo/}}

\begin{table}[htbp]
  \caption{Results of the pairwise subjective evaluation. Segment A represents the lower quality predicted by the model, and Segment B represents the higher quality. Agreement indicates the percentage of human judgments that aligned with the model predictions (i.e., preferring Segment B).}
  \label{tab:subjective}
  \begin{tabular}{lccc}
    \toprule
    \textbf{Song ID} & \textbf{Votes for Seg. A} & \textbf{Votes for Seg. B} & \textbf{Agreement} \\
    \midrule
    Song 1 & 1 & 11 & 91.7\% \\
    Song 2 & 0 & 12 & 100.0\% \\
    Song 3 & 1 & 11 & 91.7\% \\
    Song 4 & 3 & 9 & 75.0\% \\
    Song 5 & 3 & 9 & 75.0\% \\
    \midrule
    \textbf{Total} & \textbf{8} & \textbf{52} & \textbf{86.7\%} \\
  \bottomrule
\end{tabular}
\end{table}

\section{Conclusion}

In this paper, we proposed SongSQA, a two-stage framework for singing quality assessment of full-length songs. Unlike conventional clip-level paradigms, SongSQA explicitly models quality variations across different audio segments within a song and reflects their influence on the overall singing quality. To address the lack of segment-level annotations, we introduced a Segment Score Predictor trained with pseudo labels generated by a teacher model. Building upon this predictor, the Song Quality Aggregator employs unified segment embeddings and a learnable song embedding with self-attention to produce a holistic song-level quality prediction. In addition to predicting the overall quality score, SongSQA generates a temporal segment-level quality curve, providing a more fine-grained view of singing quality variations over time.
Experimental results on both datasets demonstrate the effectiveness of SongSQA for full-length song SQA, yielding up to a 13.95\% relative improvement in KTAU over the strongest baseline and consistent gains across the remaining evaluation metrics.
%




\bibliographystyle{ACM-Reference-Format}
\bibliography{sample-base}

@String{Computing = "Computing" }

@article{rozin2004feeling,
    author = {Rozin, Alexander and Rozin, Paul and Goldberg, Emily},
    title = {The Feeling of Music Past: How Listeners Remember Musical Affect},
    journal = {Music Perception},
    volume = {22},
    number = {1},
    pages = {15-39},
    year = {2004},
    month = {09},
    issn = {0730-7829},
}

@article{zoanetti2016mitigating,
  title={Mitigating the Halo Effect: Managing the Wow Factor in Music Performance Assessments},
  author={Zoanetti, Nathan and Champion, Helen},
  journal={Journal of Research in Music Performance},
  pages = {36--52},
  year={2016}
}

@article{gupta2018technical,
  title={A Technical Framework for Automatic Perceptual Evaluation of Singing Quality},
  author={Gupta, Chitralekha and Li, Haizhou and Wang, Ye},
  journal={APSIPA Transactions on Signal and Information Processing},
  volume={7},
  pages={e10},
  year={2018},
  publisher={Cambridge University Press}
}

@article{WAPNICK1997429,
title = {Expert Consensus in Solo Voice Performance Evaluation},
journal = {Journal of Voice},
volume = {11},
number = {4},
pages = {429-436},
year = {1997},
issn = {0892-1997},
author = {Joel Wapnick and Elizabeth Ekholm},
}

@inproceedings{nakano2006subjective,
  title={Subjective Evaluation of Common Singing Skills Using the Rank Ordering Method},
  author={Nakano, Tomoyasu and Goto, Masataka and Hiraga, Yuzuru},
  booktitle={Ninth International Conference on Music Perception and Cognition},
  pages={1507--1512},
  year={2006},
  organization={Citeseer}
}

@article{SUNDBERG2001176,
title = {Level and Center Frequency of the Singer's Formant},
journal = {Journal of Voice},
volume = {15},
number = {2},
pages = {176-186},
year = {2001},
issn = {0892-1997},
author = {Johan Sundberg},
}

@ARTICLE{cao2009automatic,
  author={Cao, Chuan and Li, Ming and Wu, Xiao and Suo, Hongbin and Liu, Jian and Yan, Yonghong},
  journal={IEICE Transactions on Information and Systems}, 
  title={Automatic Singing Performance Evaluation for Untrained Singers}, 
  year={2009},
  volume={E92-D},
  number={8},
  pages={1596-1600},
  ISSN={1745-1361},
  month={August},
}

@inproceedings{lal2006comparison,
  author       = {Partha Lal},
  title        = {A Comparison of Singing Evaluation Algorithms},
  booktitle    = {Ninth International Conference on Spoken Language Processing, {INTERSPEECH-ICSLP}
                  2006, Pittsburgh, PA, USA, September 17-21, 2006},
  publisher    = {{ISCA}},
  year         = {2006},
}

@ARTICLE{tsai2011automatic,
  author={Tsai, Wei-Ho and Lee, Hsin-Chieh},
  journal={IEEE Transactions on Audio, Speech, and Language Processing}, 
  title={Automatic Evaluation of Karaoke Singing Based on Pitch, Volume, and Rhythm Features}, 
  year={2012},
  volume={20},
  number={4},
  pages={1233-1243},
}

@inproceedings{molina2013fundamental,
  title={Fundamental Frequency Alignment vs. Note-Based Melodic Similarity for Singing Voice Assessment},
  author={Molina, Emilio and Barbancho, Isabel and G{\'o}mez, Emilia and Barbancho, Ana Maria and Tard{\'o}n, Lorenzo J},
  booktitle={2013 IEEE international conference on acoustics, speech and signal processing},
  pages={744--748},
  year={2013},
  organization={IEEE}
}

@inproceedings{gupta2017perceptual,
  title={Perceptual Evaluation of Singing Quality},
  author={Gupta, Chitralekha and Li, Haizhou and Wang, Ye},
  booktitle={2017 Asia-Pacific Signal and Information Processing Association Annual Summit and Conference (APSIPA ASC)},
  pages={577--586},
  year={2017},
  organization={IEEE}
}

@INPROCEEDINGS{zhang2019automatic,
  author={Zhang, Ning and Jiang, Tao and Deng, Feng and Li, Yan},
  booktitle={ICASSP 2019 - 2019 IEEE International Conference on Acoustics, Speech and Signal Processing (ICASSP)}, 
  title={Automatic Singing Evaluation without Reference Melody Using Bi-dense Neural Network}, 
  year={2019},
  volume={},
  number={},
  pages={466-470},
}

@INPROCEEDINGS{huang2020spectral,
  author={Huang, Lin and Gupta, Chitralekha and Li, Haizhou},
  booktitle={2020 Asia-Pacific Signal and Information Processing Association Annual Summit and Conference (APSIPA ASC)}, 
  title={Spectral Features and Pitch Histogram for Automatic Singing Quality Evaluation with CRNN}, 
  year={2020},
  volume={},
  number={},
  pages={492-499},
}

@INPROCEEDINGS{li2021training,
  author={Li, Jinhu and Gupta, Chitralekha and Li, Haizhou},
  booktitle={2021 Asia-Pacific Signal and Information Processing Association Annual Summit and Conference (APSIPA ASC)}, 
  title={Training Explainable Singing Quality Assessment Network with Augmented Data}, 
  year={2021},
  volume={},
  number={},
  pages={904-911},
}

@INPROCEEDINGS{cooper2022generalization,
  author={Cooper, Erica and Huang, Wen-Chin and Toda, Tomoki and Yamagishi, Junichi},
  booktitle={ICASSP 2022 - 2022 IEEE International Conference on Acoustics, Speech and Signal Processing (ICASSP)}, 
  title={Generalization Ability of MOS Prediction Networks}, 
  year={2022},
  volume={},
  number={},
  pages={8442-8446},  
}

@inproceedings{chan2023improve,
    author = {Chan, Ping-Chen and Chen, Po-Wei and Soo, Von-Wun},
    title = {Improve Singing Quality Prediction Using Self-supervised Transfer Learning and Human Perception Feedback},
    year = {2024},
    isbn = {9798400702051},
    publisher = {Association for Computing Machinery},
    address = {New York, NY, USA},
    booktitle = {Proceedings of the 5th ACM International Conference on Multimedia in Asia},
    articleno = {69},
    numpages = {7},
    location = {Tainan, Taiwan},
    series = {MMAsia '23}
}

@inproceedings{baevski2020wav2vec,
 author = {Baevski, Alexei and Zhou, Yuhao and Mohamed, Abdelrahman and Auli, Michael},
 booktitle = {Advances in Neural Information Processing Systems},
 pages = {12449--12460},
 publisher = {Curran Associates, Inc.},
 title = {wav2vec 2.0: A Framework for Self-Supervised Learning of Speech Representations},
 volume = {33},
 year = {2020}
}

@ARTICLE{hsu2021hubert,
  author={Hsu, Wei-Ning and Bolte, Benjamin and Tsai, Yao-Hung Hubert and Lakhotia, Kushal and Salakhutdinov, Ruslan and Mohamed, Abdelrahman},
  journal={IEEE/ACM Transactions on Audio, Speech, and Language Processing}, 
  title={HuBERT: Self-Supervised Speech Representation Learning by Masked Prediction of Hidden Units}, 
  year={2021},
  volume={29},
  number={},
  pages={3451-3460},
}

@INPROCEEDINGS{huang2024voicemos,
  author={Huang, Wen-Chin and Fu, Szu-Wei and Cooper, Erica and Zezario, Ryandhimas E. and Toda, Tomoki and Wang, Hsin-Min and Yamagishi, Junichi and Tsao, Yu},
  booktitle={2024 IEEE Spoken Language Technology Workshop (SLT)}, 
  title={The Voicemos Challenge 2024: Beyond Speech Quality Prediction}, 
  year={2024},
  volume={},
  number={},
  pages={803-810},
}

@article{tang2024singmos,
  title={SingMOS: An Extensive Open-Source Singing Voice Dataset for MOS Prediction},
  author={Tang, Yuxun and Shi, Jiatong and Wu, Yuning and Jin, Qin},
  journal={arXiv preprint arXiv:2406.10911},
  year={2024}
}

@article{huang2025audiomos,
  title={The AudioMOS Challenge 2025},
  author={Huang, Wen-Chin and Wang, Hui and Liu, Cheng and Wu, Yi-Chiao and Tjandra, Andros and Hsu, Wei-Ning and Cooper, Erica and Qin, Yong and Toda, Tomoki},
  journal={arXiv preprint arXiv:2509.01336},
  year={2025}
}

@INPROCEEDINGS{liu2025musiceval,
  author={Liu, Cheng and Wang, Hui and Zhao, Jinghua and Zhao, Shiwan and Bu, Hui and Xu, Xin and Zhou, Jiaming and Sun, Haoqin and Qin, Yong},
  booktitle={ICASSP 2025 - 2025 IEEE International Conference on Acoustics, Speech and Signal Processing (ICASSP)}, 
  title={MusicEval: A Generative Music Dataset with Expert Ratings for Automatic Text-to-Music Evaluation}, 
  year={2025},
  volume={},
  number={},
  pages={1-5},
}

@article{tjandra2025meta,
  title={Meta Audiobox Aesthetics: Unified Automatic Quality Assessment for Speech, Music, and Sound},
  author={Tjandra, Andros and Wu, Yi-Chiao and Guo, Baishan and Hoffman, John and Ellis, Brian and Vyas, Apoorv and Shi, Bowen and Chen, Sanyuan and Le, Matt and Zacharov, Nick and others},
  journal={arXiv preprint arXiv:2502.05139},
  year={2025}
}

@article{yao2025songeval,
  title={Songeval: A Benchmark Dataset for Song Aesthetics Evaluation},
  author={Yao, Jixun and Ma, Guobin and Xue, Huixin and Chen, Huakang and Hao, Chunbo and Jiang, Yuepeng and Liu, Haohe and Yuan, Ruibin and Xu, Jin and Xue, Wei and others},
  journal={arXiv preprint arXiv:2505.10793},
  year={2025}
}

@inproceedings{tang2025singmospro,
  title={SingMOS-Pro: An Comprehensive Benchmark for Singing Quality Assessment},
  author={Tang, Yuxun and Liu, Lan and Feng, Wenhao and Zhao, Yiwen and Han, Jionghao and Yu, Yifeng and Shi, Jiatong and Jin, Qin},
  booktitle={ICASSP 2026 - 2026 IEEE International Conference on Acoustics, Speech and Signal Processing (ICASSP)},
  year={2026},
  pages={1-5},
}

@inproceedings{wang2025singing,
  title={Singing Timbre Popularity Assessment Based on Multimodal Large Foundation Model},
  author={Wang, Zihao and Yuan, Ruibin and Geng, Ziqi and Li, Hengjia and Qu, Xingwei and Li, Xinyi and Chen, Songye and Fu, Haoying and Dannenberg, Roger B and Zhang, Kejun},
  booktitle={Proceedings of the 33rd ACM International Conference on Multimedia},
  pages={12227--12236},
  year={2025}
}

@article{bai2026reference,
title = {Reference-free singing voice MOS prediction via multi-feature fusion, with integrated feature analysis},
journal = {Applied Acoustics},
volume = {241},
pages = {110960},
year = {2026},
issn = {0003-682X},
author = {Peng Bai and Yue Zhou and Ke Gu and Meizhen Zheng and Linshujie Zheng and Yidong Chen and Xiaodong Shi},
}

@inproceedings{guo2025techsinger,
  title={TechSinger: Technique Controllable Multilingual Singing Voice Synthesis via Flow Matching},
  author={Guo, Wenxiang and Zhang, Yu and Pan, Changhao and Huang, Rongjie and Tang, Li and Li, Ruiqi and Hong, Zhiqing and Wang, Yongqi and Zhao, Zhou},
  booktitle={Proceedings of the AAAI Conference on Artificial Intelligence},
  volume={39},
  number={22},
  pages={23978--23986},
  year={2025}
}

@INPROCEEDINGS{dai2025everyone,
  author={Dai, Shuqi and Wang, Yunyun and Dannenberg, Roger B. and Jin, Zeyu},
  booktitle={ICASSP 2025 - 2025 IEEE International Conference on Acoustics, Speech and Signal Processing (ICASSP)}, 
  title={Everyone-Can-Sing: Zero-Shot Singing Voice Synthesis and Conversion with Speech Reference}, 
  year={2025},
  volume={},
  number={},
  pages={1-5},
}

@inproceedings{zhang2024stylesinger,
  title={StyleSinger: Style Transfer for Out-of-Domain Singing Voice Synthesis},
  author={Zhang, Yu and Huang, Rongjie and Li, Ruiqi and He, JinZheng and Xia, Yan and Chen, Feiyang and Duan, Xinyu and Huai, Baoxing and Zhao, Zhou},
  booktitle={Proceedings of the AAAI Conference on Artificial Intelligence},
  volume={38},
  number={17},
  pages={19597--19605},
  year={2024}
}

@article{zhang2025vevo2,
  title   = {Vevo2: A Unified and Controllable Framework for Speech and Singing Voice Generation},
  author  = {Zhang, Xueyao and Zhang, Junan and Wang, Yuancheng and Wang, Chaoren and Chen, Yuanzhe and Jia, Dongya and Chen, Zhuo and Wu, Zhizheng},
  journal = {IEEE Transactions on Audio, Speech and Language Processing},
  year    = {2026},
  pages   = {1--17}
}

@inproceedings{wang2026s2voice,
  title={{S}$^2${V}oice: Style-Aware Autoregressive Modeling with Enhanced Conditioning for Singing Style Conversion},
  author={Wang, Ziqian and Xia, Xianjun and Huang, Chuanzeng and Xie, Lei},
  booktitle={ICASSP 2026 - 2026 IEEE International Conference on Acoustics, Speech and Signal Processing (ICASSP)},
  year={2026},
  pages={1--5}
}

@INPROCEEDINGS{sha2024neural,
  author={Sha, Binzhu and Li, Xu and Wu, Zhiyong and Shan, Ying and Meng, Helen},
  booktitle={ICASSP 2024 - 2024 IEEE International Conference on Acoustics, Speech and Signal Processing (ICASSP)}, 
  title={Neural Concatenative Singing Voice Conversion: Rethinking Concatenation-Based Approach for One-Shot Singing Voice Conversion}, 
  year={2024},
  volume={},
  number={},
  pages={12577-12581},
}

@ARTICLE{zhu2025muq,
  author={Zhu, Haina and Zhou, Yizhi and Chen, Hangting and Yu, Jianwei and Ma, Ziyang and Gu, Rongzhi and Luo, Yi and Tan, Wei and Chen, Xie},
  journal={IEEE Transactions on Audio, Speech and Language Processing}, 
  title={MuQ: Self-Supervised Music Representation Learning With Mel Residual Vector Quantization}, 
  year={2025},
  volume={33},
  number={},
  pages={3653-3664},
}

@inproceedings{lv2026song,
  title     = {Song Aesthetics Evaluation with Multi-Stem Attention and Hierarchical Uncertainty Modeling},
  author    = {Lv, Yishan and Luo, Jing and Ju, Boyuan and Zhang, Yang and Wu, Xinda and Yuan, Bo and Yang, Xinyu},
  booktitle = {Proceedings of the 27th International Society for Music Information Retrieval Conference (ISMIR)},
  pages     = {1--8},
  year      = {2026}
}

@inproceedings{devlin2019bert,
    title = "{BERT}: Pre-training of Deep Bidirectional Transformers for Language Understanding",
    author = "Devlin, Jacob  and
      Chang, Ming-Wei  and
      Lee, Kenton  and
      Toutanova, Kristina",
    booktitle = "Proceedings of the 2019 Conference of the North {A}merican Chapter of the Association for Computational Linguistics: Human Language Technologies, Volume 1 (Long and Short Papers)",
    month = jun,
    year = "2019",
    address = "Minneapolis, Minnesota",
    pages = "4171--4186",
}

@inproceedings{baba2024t05,
  author       = {Kaito Baba and
                  Wataru Nakata and
                  Yuki Saito and
                  Hiroshi Saruwatari},
  title        = {The {T05} System for the voicemos challenge 2024: Transfer Learning
                  from Deep Image Classifier to Naturalness {MOS} Prediction of High-Quality
                  Synthetic Speech},
  booktitle    = {{IEEE} Spoken Language Technology Workshop, {SLT} 2024, Macao, December
                  2-5, 2024},
  pages        = {818--824},
  publisher    = {{IEEE}},
  year         = {2024},
}

@inproceedings{shi2024pitch,
  author       = {Yu{-}Fei Shi and
                  Yang Ai and
                  Ye{-}Xin Lu and
                  Hui{-}Peng Du and
                  Zhen{-}Hua Ling},
  title        = {Pitch-and-Spectrum-Aware Singing Quality Assessment with Bias Correction
                  and Model Fusion},
  booktitle    = {{IEEE} Spoken Language Technology Workshop, {SLT} 2024, Macao, December
                  2-5, 2024},
  pages        = {811--817},
  publisher    = {{IEEE}},
  year         = {2024},
}

@ARTICLE{wang2025ramp+,
  author={Wang, Hui and Zhao, Shiwan and Zheng, Xiguang and Zhou, Jiaming and Wang, Xuechen and Qin, Yong},
  journal={IEEE Transactions on Audio, Speech and Language Processing}, 
  title={RAMP+: Retrieval-Augmented MOS Prediction With Prior Knowledge Integration}, 
  year={2025},
  volume={33},
  number={},
  pages={1520-1534},
}

@inproceedings{li2024mert,
  author       = {Yizhi Li and
                  Ruibin Yuan and
                  Ge Zhang and
                  Yinghao Ma and
                  Xingran Chen and
                  Hanzhi Yin and
                  Chenghao Xiao and
                  Chenghua Lin and
                  Anton Ragni and
                  Emmanouil Benetos and
                  Norbert Gyenge and
                  Roger B. Dannenberg and
                  Ruibo Liu and
                  Wenhu Chen and
                  Gus Xia and
                  Yemin Shi and
                  Wenhao Huang and
                  Zili Wang and
                  Yike Guo and
                  Jie Fu},
  title        = {{MERT:} Acoustic Music Understanding Model with Large-Scale Self-supervised
                  Training},
  booktitle    = {The Twelfth International Conference on Learning Representations,
                  {ICLR} 2024, Vienna, Austria, May 7-11, 2024},
  year         = {2024},
}

@article{kim2021kuielab,
  title={KUIELab-MDX-Net: A Two-Stream Neural Network for Music Demixing},
  author={Kim, Minseok and Choi, Woosung and Chung, Jaehwa and Lee, Daewon and Jung, Soonyoung},
  journal={arXiv preprint arXiv:2111.12203},
  year={2021}
}

@inproceedings{jin2023order,
    author = {Jin, Xin and Zhou, Wu and Wang, Jinyu and XU, Duo and Zheng, Yongsen},
    title = {An Order-Complexity Aesthetic Assessment Model for Aesthetic-aware Music Recommendation},
    year = {2023},
    isbn = {9798400701085},
    publisher = {Association for Computing Machinery},
    address = {New York, NY, USA},
    booktitle = {Proceedings of the 31st ACM International Conference on Multimedia},
    pages = {6938–6947},
    numpages = {10},
    location = {Ottawa ON, Canada},
    series = {MM '23}
}

\end{document}